\documentclass[universe,communication,accept,pdftex,oneauthors]{Definitions/mdpi} 

\usepackage{graphicx}
\usepackage{color}
\usepackage{latexsym}

\makeatletter
\let\c@lofdepth\relax
\let\c@lotdepth\relax
\makeatother
\usepackage{subfigure}
\makeatletter
\renewcommand{\@thesubfigure}{\normalsize(\textbf{\alph {subfigure}})~}
\makeatother

\firstpage{1} 
\pubvolume{10}
\issuenum{9}
\articlenumber{378}
\pubyear{2024}
\copyrightyear{2024}
\externaleditor{Academic Editor: Firstname Lastname\vspace{-12pt}}
\datereceived{26 July 2024} 
\daterevised{2 September 2024} % Comment out if no revised date
\dateaccepted{6 September 2024} 
\datepublished{23 September 2024 } 
\hreflink{https://doi.org/} % If needed use \linebreak
\Title{{Sgr} A* Shadow Study with KTN Space Time and Investigation of NUT Charge Existence}

\TitleCitation{Sgr A* Shadow Study with KTN Space Time and Investigation of NUT Charge Existence}

\Author{{Masoumeh Ghasemi-Nodehi} \orcidA{}}
\AuthorNames{M. Ghasemi-Nodehi}

\AuthorCitation{{Ghasemi-Nodehi, M.}}
\address[1]{%
{Xinjiang} Astronomical Observatory, CAS, 150 Science1-Street, Urumqi 830011, China; mghasemin@xao.ac.cn}
\abstract{
In this paper, I investigate the existence of the NUT charge through the KTN spacetime using shadow observations of Sgr A*. I report that the range of my constraint for the NUT charge is between {$-$}0.5 and 0.5 for Schwarzschild-like and very slowly rotating KTN black holes. This range extends to 1.5 for spins up to {$-$}2 and {$-$}1.5 for spins up to 2 based on Keck observations for both 40$^\circ$ and 10$^\circ$ viewing angles. For VLTI observations, Schwarzschild-like and very slowly rotating KTN black holes are excluded for a 40$^\circ$ viewing angle, and the NUT charge is constrained to a very narrow range for a 10$^\circ$ viewing angle. I report that the possibility of having KTN naked singularities in Sgr A* is small, considering the uncertainties in the shadow size.}


\begin{document}

\def \o{\omega}
\def \n{\nabla}
\def \a{\alpha}
\def \b{\a}
\def \t{\tilde}
\def \O{\Omega}
\def \g{\gamma}
\def \s{\sigma}
\def \e{\epsilon}
\def \p{\partial}
\def \d{\Delta}
\def \x{\Xi}
\def \a{\alpha}
\def \l{\Lambda}
\def \th{\theta}
\def \r{\rho}
\def \f{\frac}

\newcommand{\bi}{\bibitem}
\newcommand{\be}{\begin{eqnarray}}
\newcommand{\ee}{\end{eqnarray}}
\newcommand{\nn}{\nonumber}
\newcommand{\bn}{\boldsymbol{\nabla}}
\newcommand{\cb}{\textcolor{red}}
\newcommand{\cbb}{\textcolor{black}}
\newcommand{\lp}{\left(}
\newcommand{\rp}{\right)}

\newcommand{\commred}{\color{red}  }

%%%%%%%%%%%%%%%%%%%%%%%%%%%%%%%%%%%%%%%%%%
% The order of the section titles is different for some journals. Please refer to the "Instructions for Authors” on the journal homepage.

\section{Introduction}\label{intro}

General Relativity predicts {that the} final product of gravitational collapse {is} Kerr black holes. These {black} holes {are} characterized by mass and spin~\cite{Kerr1,Kerr2}. The~Kerr paradigm {has been} successfully tested in weak regimes. {However, precise measurements in strong gravity regimes are still lacking.}

A test of General Relativity in {the} strong gravity regime can be executed by different methods such as Gravitational Waves~\cite{gw1,gw2}, X-ray reflection spectroscopy~\cite{iron}, {the} Continuum fitting method~\cite{con} and {the} black hole shadow~\cite{EHT1,EHT5,EHT17}. {Methods} such as X-ray reflection spectroscopy and {the} continuum fitting method are parametrically degenerate in measuring spin and deviation from General Relativity. All deformation {parameters} should vanish {to verify the} Kerr black holes~scenario.

One way to test {the} Kerr black holes paradigm is to consider {deviations} from Kerr {spacetime} {by introducing} extra parameters and {trying} to constrain the parameters. In~this work{,} I consider {the} shadow of black~holes.

The shadow of a black hole is a dark area in the observer plane. This dark area {appears} over a bright background region. The~boundary of {the} shadow depends on {the} background metric and {the} observer viewing angle in the case of {a} geometrically thick and optically thin accretion disk structure. 
So far, there are two observations {from the} Event Horizon Telescope group {of the} shadow of {a} compact object. One is {the} shadow of M87*~\cite{EHT1,EHT5} and {the other} is {the} shadow of Sgr A*~\cite{EHT17}.

In paper~\cite{ours}, {my collaborators and I} studied {the} shadow of M87* for the Kerr-Taub-NUT (KTN) {spacetime}~\cite{ml}. KTN {spacetime} has {a} Newman-Unti-Tamburino (NUT)~\cite{nut} charge in addition to Kerr parameters. {In the case of a vanishing NUT charge, the~metric reduces to the Kerr metric, and~in the case of a vanishing spin parameter, the~metric is the Taub-NUT spacetime with mass and NUT charge.}
There {are reports} in~\cite{lnbl} and also~\cite{kag,liu,cc} {about} the existence of {the} effect of {a} gravitomagnetic monopole, {also known} as NUT charge, in~{the} spectra of quasars, supernovae{,} or active galactic nuclei. Based on X-ray observations of GRO J1655-40{,} {it has been reported that this is} more consistent with KTN {spacetime} rather than Kerr {spacetime}~\cite{cbc}.

I investigate {the} possibility of {the} existence of {a} gravitomagnetic monopole{,} that is{, the} NUT charge{,} in {the} shadow of M87*. In~this paper{,} I consider {the} KTN {spacetime} again{,} but my simulation is for {the} case of {the} shadow of Sgr A*. 
In {the} M87* case{,} I found {that a} non-zero NUT charge is still compatible with {the} EHT result of M87*. I show that in prograde rotation ($a_* > 0$), $n_*$ cannot be greater than 1.1{, and} in retrograde rotation ($a_* < 0$), $n_*$ cannot be less than {$-$}1.1 %MDPI: we changed hyphen into minus sign, please confirm.
{for the} shadow of M87*. {Here, $a_*=a/M$ is the spin parameter, and~$n_*=n/M$ is the NUT charge with mass $M$.}
In addition{,} the results show that there can be {a} shadow for both cases of {the} KTN black hole and {the} KTN naked singularity. 
In this paper {on the} Sgr A* shadow, the~allowed range for {the} KTN naked singularity is narrow, so I conclude that
{the possibility of having KTN naked singularities in Sgr A* is small, considering the uncertainties in the shadow size.} The range for {the} KTN black hole is between {$-$}0.5 and 0.5 for spin zero and up to 1.5 for spin {$-$}2 and {$-$}1.5 for spin~two.

{
As for {the} shadow of naked singularities, {due to the} presence of {a} strong gravitational field{,} naked singularities can {still} cast {a} shadow. Depending on {the} geometry{,} there is {a} region {where} light {cannot} escape. Also{,} the light can be deflected and {create a shadow-like} region. The~bending or trapping can occur in a specific way. However, the~nature and {specifically the} shape and size of {the} shadow can be different from those of {a} standard black hole. The~difference {in these} distortions or asymmetry effects {can lead to} potential signature{s} {in} observation{s} to distinguish {a} naked singularity {from a} standard black hole. Even in the absence of {a} photon sphere, {due to the} strong gravitational field {resulting in} light bending{, photons can reach the observer plane}. There are {a number} of references {that} calculate {the} shadow of naked singularities. I cite some of them as~\cite{n1-2015,n2-2020,n3-2021,n4-2021,n5-2022,n6-2023}.
}

{
The significance of this study is {in} probing {deviations} from Kerr {spacetime}, quantifying {the} NUT charge or gravitomagnetic monopole, testing {the} cosmic censorship conjecture, linking observations with theory through {the} mathematically interesting theory of KTN {spacetime} {and} introducing new and exotic physics.
}

The structure of this paper is as follows. I present {the} KTN {spacetime} in Section~\ref{sec:1}. {The} shadow description is presented in Section~\ref{sec:2}. Section~\ref{sec:gheid} is devoted to constraining {the} NUT charge and {the} results of this paper. {The} conclusion is written in Section~\ref{sec:conc}.

\section{Theoretical Framework of Kerr-Taub-NUT Space~time}
\label{sec:1}
In this section, I describe the KTN space time as follows. The~metric of KTN space time can be {written as}~\cite{ml}
%MDPI: The format of equation variables (italics, bold, superscript, uppercase, lowercase, etc.) should be consistent throughout the paper (main text, tables, figures, equations). Please double check all the cases in the paper and unify them carefully.
\vspace{-6pt}
\begin{eqnarray} \nonumber
ds^2&=&-\f{\d}{p^2}(dt-A d\phi)^2+\f{p^2}{\d}dr^2+p^2 d\th^2 \nonumber \\
&+&\f{1}{p^2}\sin^2\th(adt-Bd\phi)^2
\label{metric}
\end{eqnarray}
with
\begin{eqnarray}\nonumber
\d&=&r^2-2Mr+a^2-n^2, \,\,\,\,\,\,\,\,\,  p^2=r^2+(n+a\cos\th)^2,
\\
A&=&a \sin^2\th-2n\cos\th, \,\,\,\,\,\,\,\,  B=r^2+a^2+n^2
\end{eqnarray}
where $M$ is the mass, $a_*=a/M$ is spin parameter and~$n_*=n/M$ is NUT charge. The NUT parameter is known as the gravitomagnetic monopole of a compact object as well. The~outer horizon is expressed as 
\begin{eqnarray}
 r_h &=& M(1+\sqrt{1+n_*^2-a_*^2}) .
 \label{hor}
\end{eqnarray}
{%r3
{There} %MDPI: Please confirm if this paragraph no need to add indent.
 is indication of location of singularity in KTN space time when $p^2$ vanishes~\cite{mcd},
\begin{eqnarray}
r=0 \,\,\,\,\, {\rm and} \,\,\,\, \th_s = \cos^{-1}(-n_*/a_*).
\label{sing}
\end{eqnarray}
{Expression}~\ref{sing} shows that singularity does not arise for $|n_*| > |a_*|$. Singularity is %Check intended meaning retention
 covered by horizon at $\th_s=\pi$. Depending on the numerical value of $a_*$ and $n_*$ for the KTN black hole or KTN naked singularity, singularity always arises for $|n_*| \leq |a_*|$. The~horizon, Equation~(\ref{hor}), vanishes for $|a_*| > \sqrt{1+n_*^2}$. This leads to KTN naked singularity. Meanwhile, for $|n_*| \leq |a_*| \leq \sqrt{1+n_*^2}$, the~KTN black hole with (covered) singularity is obtained. 
}
One can refer to paper~\cite{ours} for more information about the metric of space time. 
\section{Methods and Description of Shadow Boundary of a Compact~Object}
\label{sec:2}
{The} image of {a} black hole, {also called the} black hole shadow, is a dark area over a brighter region. Capture, scatter to infinity{,} and {the} critical curve are the trajectories of {photons} approaching {the} black hole from distant radii around the black hole. {The} critical curve is the separation {between} capture and scatter to infinity trajectories. In~{the} case of {a} nearly tangential to circular orbit, {the} 3-momentum of {the} light ray is unstable and {orbits} around {the} black hole several times{,} which {creates a} brighter area around {the} darker region. The~2D dark region in {the} observer's sky plane is {the} shadow of {the} black hole. To~obtain the shadow for {a} specific {spacetime} geometry, one should solve {the} geodesic equations. In~this paper{,} I solve {these} equations numerically using {the} ray tracing technique. Solving {the} equations {is achieved} with a class of the Runge--Kutta--Nystrom method with adaptive step size {and} error control~\cite{Lund}. {Ray tracing images the observer's perception of a distant object or portion of the sky. Each observed light ray is traced back to the object. Ray tracing studies the distortion or appearance of a background distant star by local gravitational~fields.

The Runge--Kutta--Nystrom method is a numerical technique for second-order ordinary differential equations. There is no need to convert them into a system of first-order equations. I start from equations like
\begin{eqnarray}
\frac{\textrm{d}^2 y}{\textrm{d}\tau^2}  = f(\tau, y, \frac{\textrm{d} y}{\textrm{d}\tau})\,.
\end{eqnarray}

As geodesic equations are second-order differential equations,
I can use the Runge--Kutta--Nystrom method to solve them numerically.
In the case of geodesic equations

\begin{eqnarray}
\frac{\textrm{d}^2 x^{\mu}}{\textrm{d}\tau^2} + \Gamma^{\mu}_{\alpha \beta} \frac{\textrm{d}x^{\alpha}}{\textrm{d}\tau} \frac{\textrm{d}x^{\beta}}{\textrm{d}\tau} = 0\,,
\end{eqnarray}

I have $y = x^{\mu}(\tau)$. The~Runge--Kutta--Nystrom method updates the position of $x^{\mu}$ and $\frac{\textrm{d}x^{\mu}}{\textrm{d}\tau}$. It uses a specific iterative scheme to perform this update. As~for adaptive step sizes{,} I have control {over} errors in each step{,} ensuring the solution remains accurate by controlling the errors dynamically and computing with high precision.}

{I start from the observer plane.}
{The} observer's sky plane in my setup is located at the {Cartesian} coordinate $\left( x' , y', z' \right)$ and {the} compact object is at $\left( x , y, z \right)$.
%as the Figure~1 and 3 of paper~\cite{myshadow1,myshadow2} respectively. 
{The} distance of {the} observer is $D$ and the viewing angle is $i$. ${ k_0}=(k_0^t,k_0^r,k_0^\theta,k_0^\phi)$ is {the} photon 4-momentum perpendicular to {the} observer plane {from which} I fire {the} photon. This observer plane is a grid of fired photons and the ray is traced back. The~initial conditions are as~follows.

{The transformation from image plane $ x' , y', z' $ to compact object $x , y, z$ is
\be \label{coor1}
x &=& D \sin (i)-y' \cos (i)+z' \sin (i)\,, \nonumber \\
y &=& x'\,, \nonumber \\
z &=& D \cos (i)+y' \sin (i)+z' \cos (i) \,.
\ee

\textls[+30]{The coordinate system reduces to a spherical coordinate far from the compact object \mbox{with transformation.}}

\be \label{coor2}
r &=& \sqrt{x^2+y^2+z^2}  \,,\nonumber \\
\theta &=& \arccos \lp \frac{z}{r} \rp \,, \nonumber \\
\phi &=& \arctan\left(\frac{y}{x}\right)\,.
\ee

Considering photon initial position at $\lp x'_0, y'_0, 0 \rp$ with 3-momentum ${ k_0} = -k_0 z'$ perpendicular to the image plane, the~photon initial conditions are obtained %Check intended meaning retention
by substituting Equation~(\ref{coor1}) into Equation~(\ref{coor2}) at $\lp x'_0, y'_0, 0 \rp$
},
\vspace{-3pt}
\be
\label{eq-1}
t_0 &=& 0 \, , \nonumber\\
r_0 &=& \sqrt{x'^2_0 + y'^2_0 + D^2} \, , \nonumber\\
\theta_0 &=& \arccos \frac{y'_0 \sin i + D \cos i}{\sqrt{x'^2_0 + y'^2_0 + D^2}} \, , \nonumber\\
\phi_0 &=& \arctan \frac{x'_0}{D \sin i - y'_0 \cos i} \, .
\ee

The initial condition of the 4-momentum is
\vspace{-3pt}
\be
\label{eq-2}
k^r_0 &=& - \frac{D}{\sqrt{x'^2_0 + y'^2_0 + D^2}} |k_0| \, , \nonumber\\
k^\theta_0 &=& \frac{\cos i - D \frac{y'_0 \sin i + D 
\cos i}{x'^2_0 + y'^2_0 + D^2}}{\sqrt{x'^2_0 + (D \sin i - y'_0 \cos i)^2}} |k_0| \, , \nonumber\\
k^\phi_0 &=& \frac{x'_0 \sin i}{x'^2_0 + (D \sin i - y'_0 \cos i)^2}|k_0| \, , \nonumber\\
k^t_0 &=& \sqrt{\left(k^r_0\right)^2 + r^2_0  \left(k^\theta_0\right)^2
+ r_0^2 \sin^2\theta_0  (k^\phi_0)^2} \, 
\ee
where  $g_{\mu\nu} k^{\mu} k^{\nu} = 0$ is used to calculate $k^t_0$.

To {define the} boundary of {the} shadow, {I use the following definition: $\rho(x',y') = 1$ for inside the shadow and $\rho(x',y') = 0$ for outside the shadow.}
\be
x'_{\rm center} &=& \frac{\int \int \rho(x',y') x' dx' dy'}{\int \int \rho(x',y') dx' dy'} \,  \nonumber\\
y'_{\rm center} &=& \frac{\int \int \rho(x',y') y' dx' dy'}{\int \int \rho(x',y') dx' dy'} \, ,
\ee

\begin{figure}[H]
%
  %\begin{center}
  %\resizebox{0.5\textwidth}{!}{
  \includegraphics[width=2in]{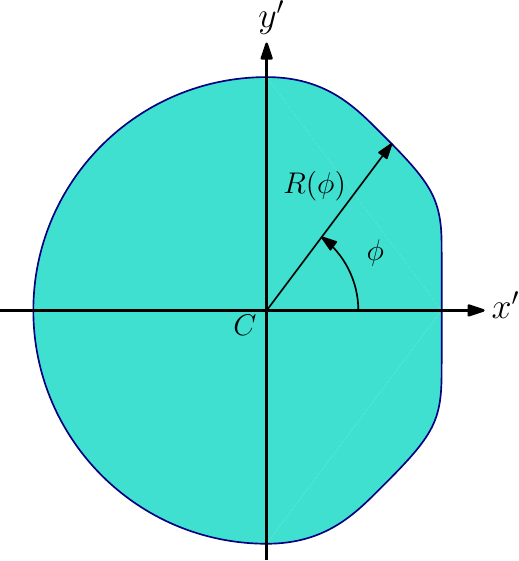}
  %  %
  %\end{center}
%
\caption{\label{shbou} {I first measure center of shadow as discussed in the text. $\phi = 0$ is from shorter segment of symmetry axis of shadow which is $x'$. Starting from $\phi = 0$, $R \left( \phi \right)$ defines each point of boundary of shadow from center of~shadow.}}
\end{figure}

$\phi = 0$ is {the} shorter segment of {the} symmetry axis of {the} shadow{,} which is $x'$. Starting from $\phi = 0$, $R \left( \phi \right)$ defines each point of {the} boundary of {the} shadow from {the} center of {the} shadow. {Please see Figure~\ref{shbou} for shadow boundary definitions. Also refer to
~\cite{myshadow1,myshadow2} for shadow description.}

\section{Constraining the NUT Charges  in Sgr A*}
\label{sec:gheid}

EHT collaborations~\cite{EHT17} {characterize the} deviation from {the} shadow size of {the} Schwarzschild {black hole} of Sgr A* as
\be\label{delta}
\delta_{\rm metric} = \frac{\tilde d_{\rm metric}}{6\sqrt{3}} -1
\ee
where $\tilde d_{\rm metric}$ is {the} shadow size{,} that is{,} the diameter between two points of {the} boundary of {the} shadow along {the} $x'$ axis. {The} size of {the} Schwarzschild shadow is $6\sqrt{3}$; it is circular. %Check intended meaning retention
The authors report that the $\delta$ values for Sgr A* {are} as follows~\cite{EHT17}:
\be\label{EHT}
{\rm Keck:} &&\delta = -0.04^{+0.09}_{-0.10} \nonumber \\
{\rm VLTI:} &&\delta = -0.08^{+0.09}_{-0.09}
\ee

In this paper, I use these values to constrain {the} NUT~charge.

\begin{figure}[H]
%\begin{center}
  \subfigure[Keck]{\includegraphics[width=2.5in,angle=0]{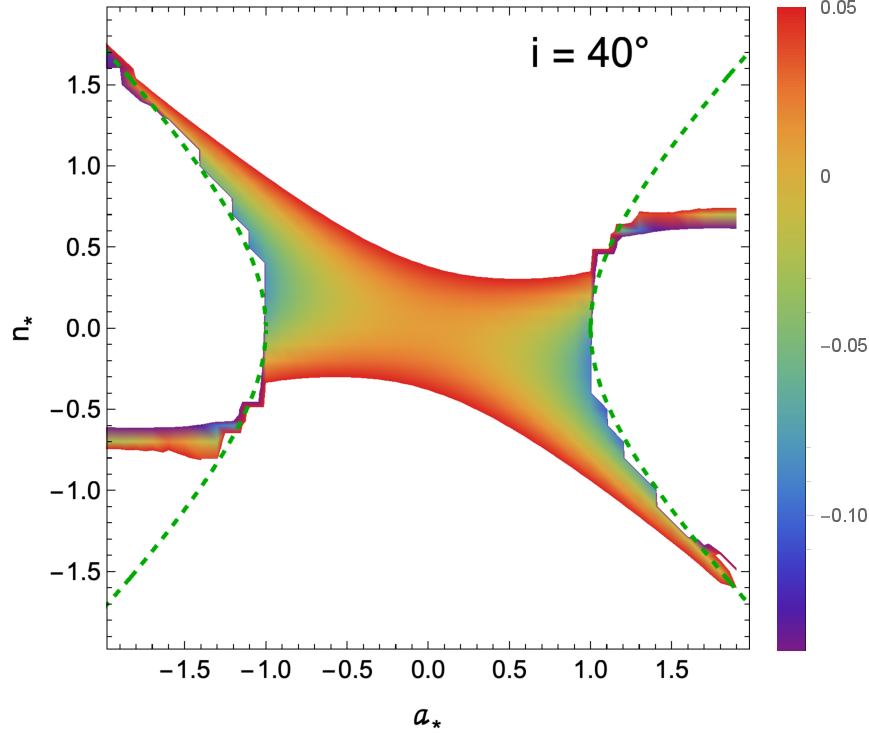}}
\hspace{0.05\textwidth}
\subfigure[VLTI]{\includegraphics[width=2.5in,angle=0]{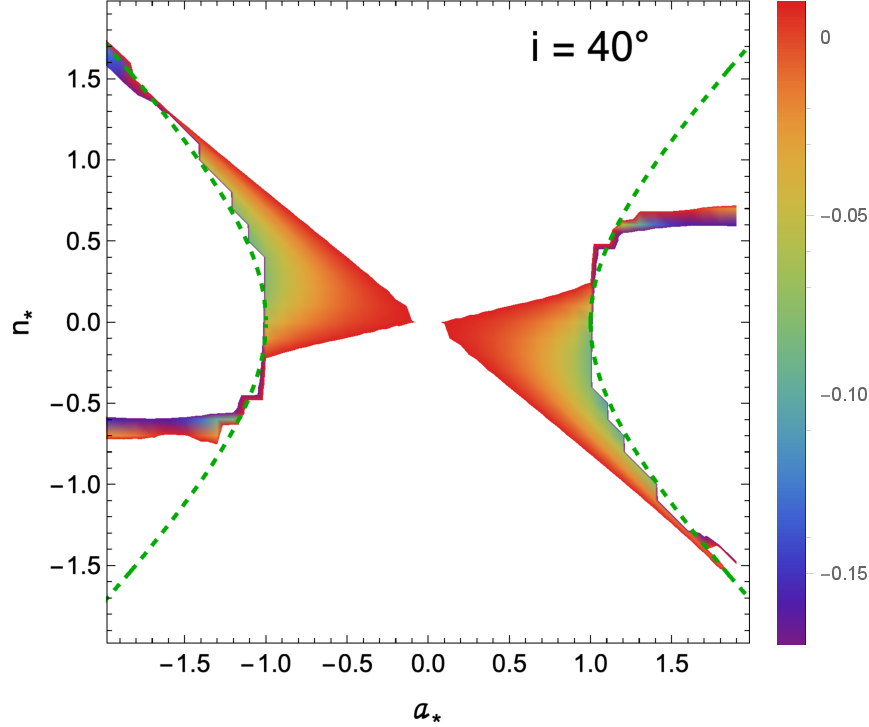}}
\hspace{0.05\textwidth}
\caption{\label{fig:40} The colored regions show deviation from circular shadow of Schwarzschild black hole that is $\delta_{\rm metric}$ in in Equation~(\ref{delta}). The~range is according to EHT observation of Keck, (\textbf{a}), and~VLTI, (\textbf{b}), as~Equation~(\ref{EHT}). The~viewing angle is 40$^\circ$ in these plots. The~central region within the two dashed green lines is KTN black holes and the outer part is KTN naked~singularities. }
%\end{center}   
\end{figure}

\begin{figure}[H]
\begin{center}
  \subfigure[Keck]{\includegraphics[width=2.5in,angle=0]{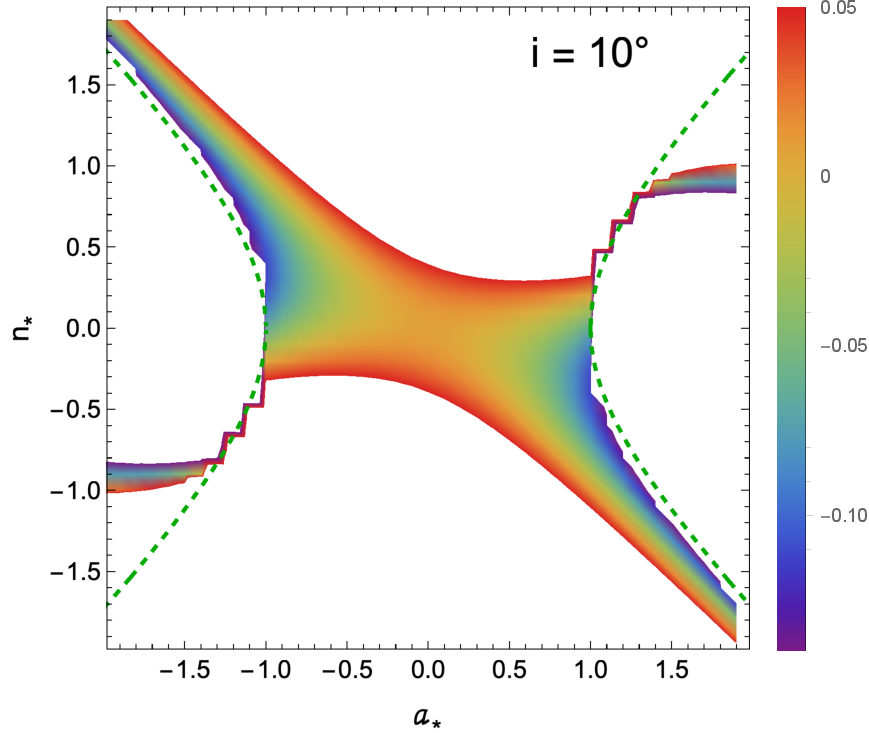}}
\hspace{0.05\textwidth}
\subfigure[VLTI]{\includegraphics[width=2.5in,angle=0]{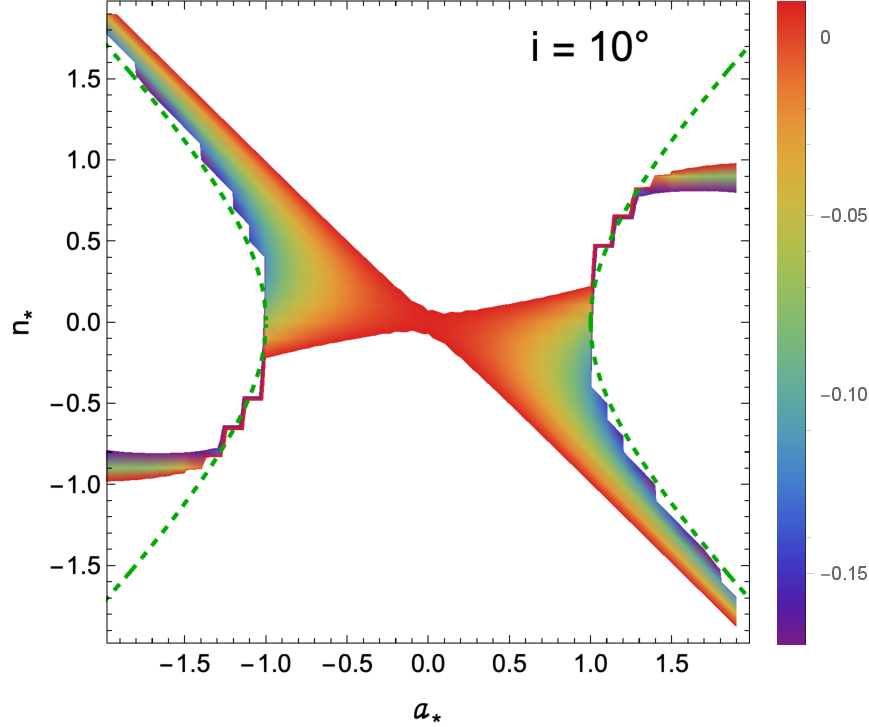}}
\hspace{0.05\textwidth}
\caption{\label{fig:10} The colored regions show deviation from circular shadow of Schwarzschild black hole that is $\delta_{\rm metric}$ in Equation~(\ref{delta}). The~range is according to EHT observation of Keck, (a), and~VLTI, (b), as~in Equation~(\ref{EHT}).  The~viewing angle is 10$^\circ$ in these plots. The~central region within the two dashed green lines is KTN black holes and the outer part is KTN naked~singularities. }
\end{center}   
\end{figure}

I calculate $\delta_{\rm metric}$ of Equation~(\ref{delta}) for each $a_*$ and $n_*$. I consider {a} viewing angle {of} 40$^\circ$ and 10$^\circ$. I plot both values of {the} reported $\delta$ of Keck and VLTI observations. \mbox{Figures~\ref{fig:40} and \ref{fig:10}} show my results. I limit the color region to {the} value reported in {Equation}~\ref{EHT} as {the} EHT collaboration stated~\cite{EHT17}.
\vspace{-6pt}
\unskip

As can be seen in {plots} {Figures}~\ref{fig:40}a and~\ref{fig:10}a for Keck observations, {the} NUT charge can vary from about {$-$}0.5 to about 0.5 for {the} central region{,} that is{,} Schwarzschild and very slowly rotating KTN black holes. The~colored region is extended to {a} NUT charge {of} 1.5 for a negative spin up to {$-$}2 and {$-$}1.5 for a positive spin up to 2 for KTN black~holes.
%MDPI: We added ``Figure'' and revise format please  confirm.

As can be seen in {plot} {Figure}~\ref{fig:40}b for VLTI observations, the~central region{,} that is, for Schwarzschild-like and very slowly rotating KTN black holes{,} {is} excluded. However{,} for {plot} {Figure}~\ref{fig:10}b for VLTI observations, the~middle part is not excluded{,} but the constrained range for {the} NUT charge is narrow and close to the zero value. The~colored region is extended to {a} NUT charge {of} 1.5 for a negative spin up to {$-$}2 and {$-$}1.5 for a positive spin up to 2 for KTN black~holes.
 
{In Figures~\ref{fig:0.240}--\ref{fig:0.810}{,} I fix some spin values as examples in order to see better ranges for {the} NUT charge. The~red dashed lines in all plots show the limits from {the} EHT observation of Sgr A*, reported in {Equation} (\ref{EHT}) for $\delta$. As~discussed above, for~slowly rotating KTN black {holes}, such as {the} plots in Figures~\ref{fig:0.240} and~\ref{fig:0.210} with {a} spin {of} 0.2, the~constraints are narrower and for VLTI observations {are} nearly zero. For~larger spin values{,} such as 0.8 in Figures~\ref{fig:0.840} and~\ref{fig:0.810}{,} it spans {a} larger range as shown in the plots. 
All plots include {a} zero value for {he} NUT charge. {A} slowly rotating KTN black hole in the sample plots, \mbox{Figures~\ref{fig:0.240} and~\ref{fig:0.210}}, does not contain negative values for $\delta${,} but {a} stronger gravitational field {with} larger spin values in {the} plots, Figures~\ref{fig:0.840} and~\ref{fig:0.810}{,} introduces negative values for $\delta${,} {causing the} shape {to} become distorted {compared} with {the} shape of {Schwarzschild black holes}.

\begin{figure}[H]
\begin{center}
  \subfigure[Keck]{\includegraphics[width=2.5in,angle=0]{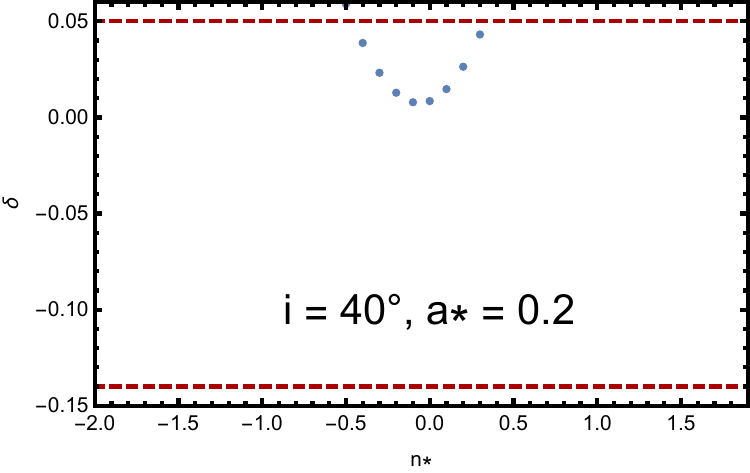}}
\hspace{0.05\textwidth}
\subfigure[VLTI]{\includegraphics[width=2.5in,angle=0]{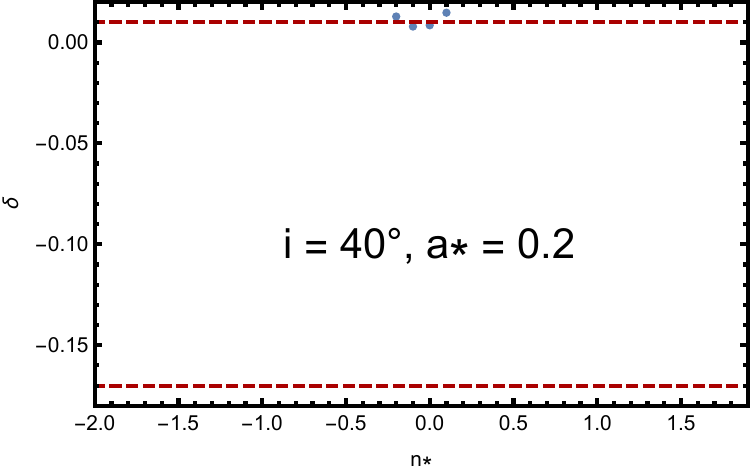}}
\hspace{0.05\textwidth}
\caption{\label{fig:0.240} Plot for fixed value of spin, 0.2, and~viewing angle 40$^\circ$. The~red dashed line is observational limits from EHT~collaborations.}
\end{center}   
\end{figure}
\unskip

\begin{figure}[H]
\begin{center}
  \subfigure[Keck]{\includegraphics[width=2.5in,angle=0]{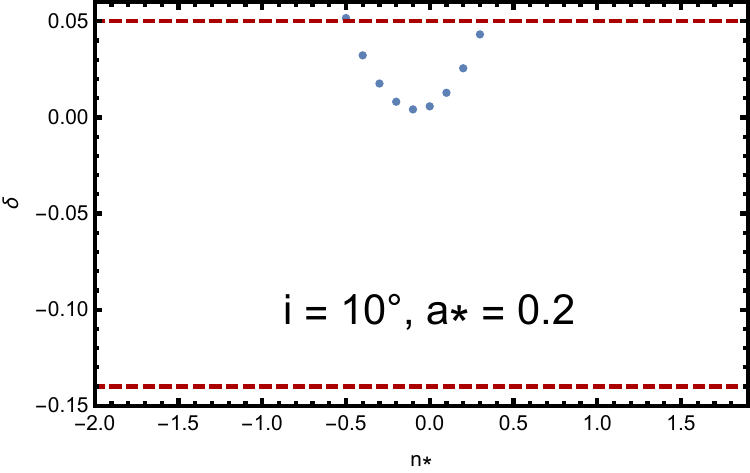}}
\hspace{0.05\textwidth}
\subfigure[VLTI]{\includegraphics[width=2.5in,angle=0]{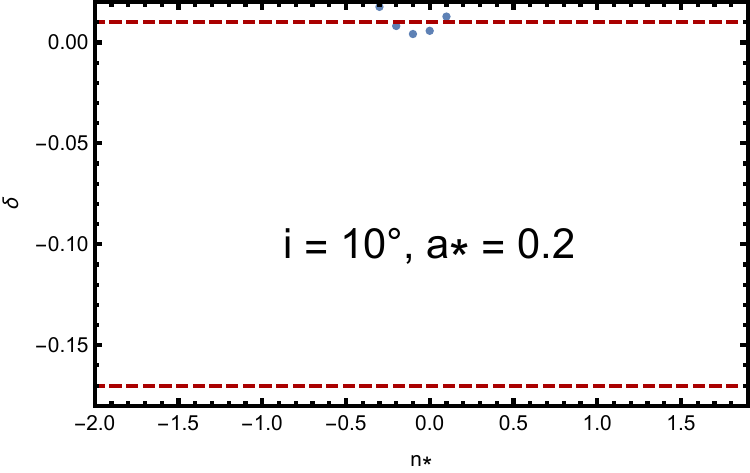}}
\hspace{0.05\textwidth}
\caption{\label{fig:0.210} Plot for fixed value of spin, 0.2, and~viewing angle 10$^\circ$. The~red dashed line is observational limits from EHT~collaborations. }
\end{center}   
\end{figure}
\vspace{-18pt}

\begin{figure}[H]
\begin{center}
  \subfigure[Keck]{\includegraphics[width=2.5in,angle=0]{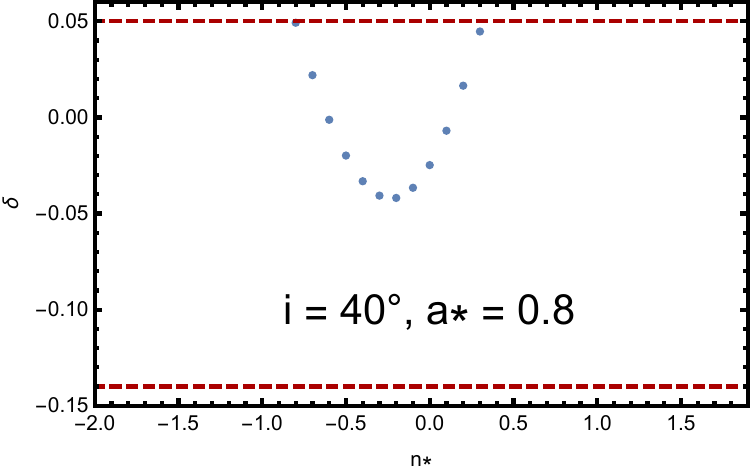}}
\hspace{0.05\textwidth}
\subfigure[VLTI]{\includegraphics[width=2.5in,angle=0]{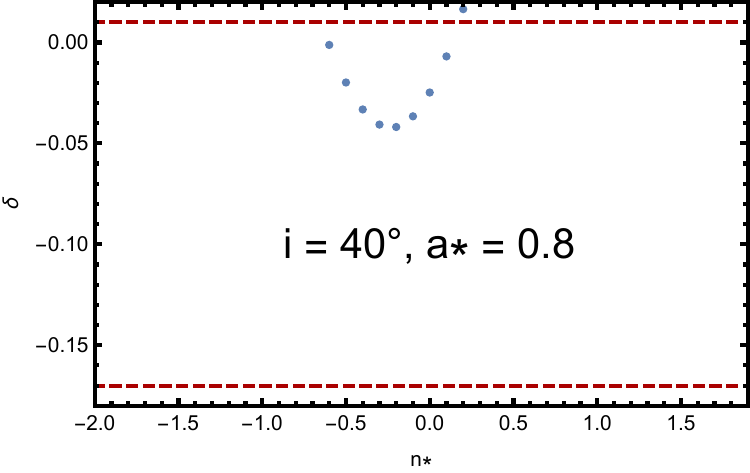}}
\hspace{0.05\textwidth}
\caption{\label{fig:0.840} Plot for fixed value of spin, 0.8, and~viewing angle 40$^\circ$. The~red dashed line is observational limits from EHT~collaborations.}
\end{center}   
\end{figure}
\vspace{-18pt}
\begin{figure}[H]
\begin{center}
  \subfigure[Keck]{\includegraphics[width=2.5in,angle=0]{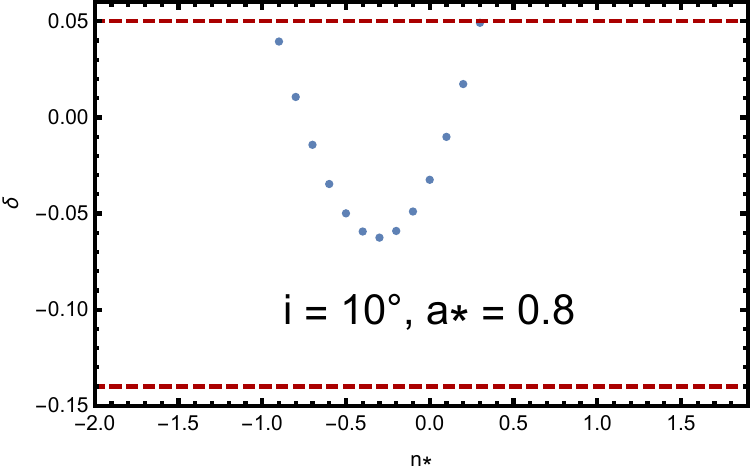}}
\hspace{0.05\textwidth}
\subfigure[VLTI]{\includegraphics[width=2.5in,angle=0]{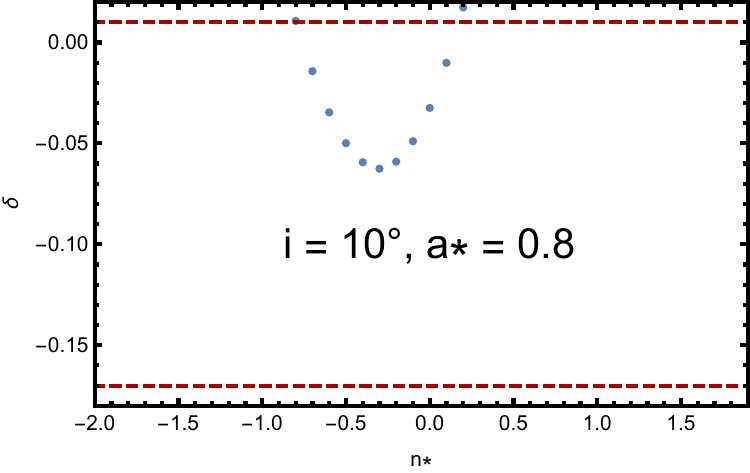}}
\hspace{0.05\textwidth}
\caption{\label{fig:0.810} Plot for fixed value of spin, 0.8, and~viewing angle 10$^\circ$. The~red dashed line is observational limits from EHT~collaborations. }
\end{center}   
\end{figure}
}

As the colored region for KTN naked singularities is narrow and considering uncertainties, I conclude that  the possibility for Sgr A* as a KTN naked singularities is small for both cases of Keck and VLTI for both 40$^\circ$ and 10$^\circ$ viewing angles.  
{Qualitatively} speaking, according to Figure~\ref{fig:10}, for~a fixed spin value within the naked singularity region, the~color distribution is symmetric, with~the middle part showing a more bluish or greenish hue, while moving towards higher or lower NUT charges shifts the color towards red. This indicates that the middle part corresponds to more negative \mbox{$\delta$ values}, transitioning to positive values as the color becomes redder. This implies that the shadow shape is becoming oblate as it transitions to a reddish color. However, when the viewing angle is increased to 40$^\circ$ in Figure~\ref{fig:40}, the~colors become mixed, and~no specific trend is observed. Nevertheless, qualitatively speaking, the~dominant color in the Keck observation is red, and~possibly yellow (positive values), with~a smaller blue region. For~VLTI, the~blue color (negative values) is more prominent. In~all cases depicted in \mbox{Figures~\ref{fig:40} and~\ref{fig:10}}, the~NUT charge appears constant with increasing spin values ($|a_*|$) in the naked \mbox{singularity region. }
 %MDPI: We removed the bold. Please confirm this revision.

\section{Conclusions and~Discussion}
\label{sec:conc}

{The KTN space time is generalization of the Kerr space time as a rotating black hole in General Relativity. The~KTN space time introduces additional parameter, the NUT charge. As~key features of the KTN space time, it is rotating like the Kerr case with mass and angular momentum. The~additional parameter, the NUT charge or a related gravitomagnetic monopole, provides a non-trivial topology and alters the causal structure. Like the Kerr case, the off-diagonal component in the metric leads to a frame-dragging signature, but the KTN case is more complex because of the presence of the NUT charge. %Check intended meaning retention
	As~for limitations, the~physical interpretation of the NUT charge is not clear in nature. It is kind of a dual mass or a gravitational analog to a magnetic monopole that there is no direct observational evidence for its existence so far. The~KTN space time is mathematically interesting, but it physical implications and relevance to astrophysics is an open question.} %Check intended meaning retention

In this paper{,} I {investigate the} existence of {the} NUT charge {in the} shadow of Sgr A* reported by {the} EHT collaborations. I consider {the} metric of KTN {spacetime} in this study. I calculate {the} shadow of Sgr A* with KTN {spacetime,} including {the} NUT charge. Then, I constrain {my results} with {the} reported values of deviations from {the} Schwarzschild shadow size from {the} EHT collaborations. In~a previous paper{,} we \mbox{calculate{d}}, {the} shadow of the same {spacetime} for M87*. We show that M87* can be both {a} KTN black hole and {a} KTN naked singularity. We report {that the} upper limit for {the} NUT charge {is} not greater {than} 1.1 for a positive spin. The~lower limit {reports that the} NUT charge cannot be less {than} -1.1 for negative spin values. Also{,} {a} KTN naked singularity can still be compatible with M87* observations.

In this work, the~ranges for constraining {the} NUT {charge} from the Keck observation of Sgr A* can be {a} maximum {of} 0.5 for Schwarzschild-like and very slowly rotating KTN black holes,
and {extended to} 1.5 for negative spin values up to {$-$}2. The~minimum value for {the} NUT charge for Keck observations can be {$-$}0.5 for 
Schwarzschild-like and very slowly rotating KTN black holes, and~{extended to} {$-$}1.5 for positive spin values up \mbox{to 2.} I consider two values of 40$^\circ$ and 10$^\circ$ for viewing angles. For~VLTI observations, the~Schwarzschild-like and very slowly rotating KTN black holes are excluded for 40$^\circ$ and {the range is} very narrow in {the} NUT charge for 10$^\circ$. The~colored region is extended to 1.5 for {the} NUT charge for negative spin values up to {$-$}2 and {$-$}1.5 for positive spin values up to 2. For~all four plots I report{,} there is {a} small possibility of indications of KTN naked singularities considering~uncertainties.

{
The main significance of {the} result is testing General Relativity and exploring alternative gravity theories, deviations from {the} Kerr solution{,} and exploring {the} possibility of exotic physics. Constraining {the} NUT charge{,} which would be interpreted as {the} gravitational analog of {a} magnetic monopole through observations{,} could {lead to} certain extensions and modifications of standard General Relativity. Confirming {the} existence of {the} NUT charge has {implications} for {the} type of objects that can exist, {the} end state {of} gravitational collapse and black hole formation{and} Galactic dynamics.
}

{
In addition to {the} possible constraining of {exotic solutions} and {the} implications of {a} gravitational monopole, my work {supports} the cosmic censorship conjecture (singularities should be hidden {within the} event horizon) by reducing {the} possibility of {the} existence of naked singularities in {the} shadow of Sgr A*. Naked singularities are a theoretical challenge in General Relativity.
}

{
In general, {the} possibility of singularities with other observations is reduced. As~one example, one can study X-ray data through {the} iron line method to probe strong gravity regimes. X-ray reflection spectroscopy{,} {also known} as {the} iron line method{,} studies {the} prominent feature of {the} iron $K\alpha$ line in X-ray reflection spectra. In~the inner region of {the} accretion disk, close to {the} compact object in {the} presence of {a} strong gravitational field, the~line is broadened and asymmetric. Currently, this method is used for measuring black hole spin considering {the} Kerr {spacetime}. It is also used to test {deviations} from {the} Kerr case. In~paper ~\cite{nx}, the~authors show that {the} iron line of naked singularities {is} different from that of {a} black hole. For~instance, the~line introduces three peaks instead of \mbox{two peaks}. One can also simulate {the} KTN naked singularity iron line and see the~differences.

Regarding {the} test of General Relativity and black holes through X-ray observations{,} there are some problems as well. These observations suffer from {parametric degeneracy} problems with current {facilities}. That means that the same observations can be fitted by considering {a} spacetime deviated from {the} Kerr case. Shadow observations {and} the narrow range of {the} NUT charge in this study can help to better understand {alternatives} to {the} Kerr~paradigm.

Moreover, in~another recent manuscript~\cite{kia}, the~authors consider some naked singular {spacetimes} and {study them using the} shadow of Sgr A* and M87*. They show {that a} broad class of naked singularities {is} excluded.
} 

Regarding the viewing angle, one can consider different angles {as there are} no strong {constraints} from observation {on} the viewing angle. The~EHT group only report values below 50$^\circ$, so I consider 40$^\circ$ and 10$^\circ$ as examples. There is also some reported range from infrared, radio, and other observations; it is mostly between 30$^\circ$ and 60$^\circ$. One can also consider higher values, but as {they are} not reported by observations, do not consider {them} here.

A crucial point is understanding the impact of plasma on {the} shadow; for example, {} blob of plasma can affect {it} as {a} Doppler shift on {the} shadow image due to {the} plasma’s high velocity and the {resulting} relativistic Doppler effect. {Additionally,} it can distort the appearance of {the} black hole. Observationally, {the} brightness of {the} photon ring {caused by} moving/orbiting plasma {causes} variability in time for {the} shadow image. GRMHD simulations can model the plasma and its behavior. The~Doppler shift and beaming signature can be studied. The~brightness and spectrum of {the} accretion disk and jet can lead {to an} asymmetric structure in {these} studies,  and the {resulting} photon ring appearance would be asymmetric as~well.

%%%%%%%%%%%%%%%%%%%%%%%%%%%%%%%%%%%%%%%%%%%%%%%%%%%%%%%%%

%%%%%%%%%%%%%%%%%%%%%%%%%%%%%%%

%%%%%%%%%%%%%%%%%%%%%%%%%%%%%%%%%%%%%%%%%%
\vspace{6pt} 

%%%%%%%%%%%%%%%%%%%%%%%%%%%%%%%%%%%%%%%%%%
%% optional
%\supplementary{The following supporting information can be downloaded at:  \linksupplementary{s1}, Figure S1: title; Table S1: title; Video S1: title.}


%\begin{tabularx}{\textwidth}{lll}
%\toprule
%\textbf{Name} & \textbf{Type} & \textbf{Description} \\
%\midrule
%S1 & Python script (.py) & Script of python source code used in XX \\
%S2 & Text (.txt) & Script of modelling code used to make Figure X \\
%S3 & Text (.txt) & Raw data from experiment X \\
%S4 & Video (.mp4) & Video demonstrating the hardware in use \\
%\bottomrule
%\end{tabularx}
%}

%%%%%%%%%%%%%%%%%%%%%%%%%%%%%%%%%%%%%%%%%%

\funding{{This paper} is supported by the CAS Talent Program and the Xinjiang Tianchi Talent~Program}
%MDPI: we removed author contributions part because there is only one author, please confirm. 

%\institutionalreview{Not applicable.}

%\informedconsent{Not applicable}

\dataavailability{ The raw data supporting the conclusions of this article will be made available by the authors on request.} 

% Only for journal Nursing Reports
%\publicinvolvement{Please describe how the public (patients, consumers, carers) were involved in the research. Consider reporting against the GRIPP2 (Guidance for Reporting Involvement of Patients and the Public) checklist. If the public were not involved in any aspect of the research add: ``No public involvement in any aspect of this research''.}

% Only for journal Nursing Reports
%\guidelinesstandards{Please add a statement indicating which reporting guideline was used when drafting the report. For example, ``This manuscript was drafted against the XXX (the full name of reporting guidelines and citation) for XXX (type of research) research''. A complete list of reporting guidelines can be accessed via the equator network: \url{https://www.equator-network.org/}.}

% Only for journal Nursing Reports
%\useofartificialintelligence{Please describe in detail any and all uses of artificial intelligence (AI) or AI-assisted tools used in the preparation of the manuscript. This may include, but is not limited to, language translation, language editing and grammar, or generating text. Alternatively, please state that “AI or AI-assisted tools were not used in drafting any aspect of this manuscript”.}

\acknowledgments{I acknowledge the support from the CAS Talent Program and the Xinjiang Tianchi Talent Program. I also thank Youjun Lu and {{Golshan~Ejlali}} for fruitful discussions. Discussion with Chandrachur Chakraborty at early stage of work is appreciated as well. I would like to express my sincere gratitude to the referees for their valuable comments and suggestions, which have greatly improved the quality of this manuscript.}

\conflictsofinterest{The author declares no conflicts of~interest.} 

%%%%%%%%%%%%%%%%%%%%%%%%%%%%%%%%%%%%%%%%%%
%% Optional

%% Only for journal Encyclopedia
%\entrylink{The Link to this entry published on the encyclopedia platform.}

\abbreviations{Abbreviations}{
The following abbreviations are used in this manuscript:\\

\noindent 
\begin{tabular}{@{}ll}
KTN & Kerr-Taub-NUT\\
NUT & Newman-Unti-Tamburino\\
EHT & Event Horizon Telescope
\end{tabular}
}

\newpage
%%%%%%%%%%%%%%%%%%%%%%%%%%%%%%%%%%%%%%%%%%
%% Optional
%%%%%%%%%%%%%%%%%%%%%%%%%%%%%%%%%%%%%%%%%%
\begin{adjustwidth}{-\extralength}{0cm}

\reftitle{References}

\PublishersNote{}
\end{adjustwidth}
\end{document}